\documentclass[12pt]{article}
\usepackage[utf8]{inputenc}

\usepackage[style=authoryear]{biblatex}
\usepackage{times}

\usepackage{ amssymb }
\usepackage{graphicx}
\usepackage{tikz}
\usepackage{tikz}
	\usetikzlibrary{shapes,arrows, chains, positioning, shadings, calc, decorations}

\usepackage{float}
\usepackage{geometry}
\date{\today}

\newcommand{\MJMA}{$\mathrm{M_{JMA}}$}
\newcommand{\ML}{$\mathrm{M_{L}}$}
\newcommand{\MV}{$\mathrm{M_{V}}$}
\newcommand{\MS}{$\mathrm{M_{S}}$}
\newcommand{\mb}{$\mathrm{m_{b}}$}
\newcommand{\Mw}{$\mathrm{M_{w}}$}
\newcommand{\EQ}{\texttt{EQ}\ }
\newcommand{\nEQ}{\texttt{nEQ}\ }

\title{Testing the Potential of Deep Learning in Earthquake Forecasting}
\author{Jonas Köhler,$^{1,2, \dag}$ Wei Li,$^{1}$ Johannes Faber,$^{1,3}$\\
Georg Rümpker,$^{1,2}$ Nishtha Srivastava$^{1,2,\ast}$\\
\\
\footnotesize{$^{1}$Frankfurt Institute of Advanced Studies, Ruth-Moufang-Str.~1, 60438 Frankfurt am Main, Germany}\\
\footnotesize{$^{2}$Institute of Geosciences, Goethe-University Frankfurt, 60438 Frankfurt am Main, Germany}\\
\footnotesize{$^{3}$Institute for Theoretical Physics, Goethe Universität, 60438 Frankfurt am Main, German}\\
\\
\footnotesize{Corresponding Authors; E-mail:  $^\dag$jkoehler@fias.uni-frankfurt.de, $^\ast$srivastava@fias.uni-frankfurt.de.}
}

\date{}

\begin{document}

\maketitle
\section*{Abstract}
Reliable earthquake forecasting methods have long been sought after, and so the rise of modern data science techniques raises a new question: does deep learning have the potential to learn this pattern? 
In this study, we leverage the large amount of earthquakes reported via good seismic station coverage in the subduction zone of Japan.  We pose earthquake forecasting as a classification problem and train a Deep Learning Network to decide, whether a timeseries of length $\geq$ 2 years will end in an earthquake on the following day with magnitude $\geq$ 5 or not.
 
Our method is based on spatiotemporal b value data, on which we train an autoencoder to learn the normal seismic behaviour. We then take the pixel by pixel reconstruction error as input for a Convolutional Dilated Network classifier, whose model output could serve for earthquake forecasting. We develop a special progressive training method for this model to mimic real life use. The trained network is then evaluated over the actual dataseries of Japan from 2002 to 2020 to simulate a real life application scenario. The overall accuracy of the model is $72.3\%$. The accuracy of this classification is significantly above the baseline and can likely be improved with more data in the future.

\section{Introduction}


Earthquakes are some of the most destructive and unpredictable natural phenomena on earth, and their forecasting has posed a significant problem for seismologists. The ongoing extensive data collection coupled with a growing understanding of seismic processes could have the potential to improve this in the future.

Previous forecasting attempts \parencite{Smith1981, Main1989, Smyth2011, Gulia2019} have frequently used the empirical Gutenberg Richter Law \parencite{Gutenberg1944}
\begin{equation}
    \log_{10} N = a - b M
\end{equation}  which gives a logarithmic relation between the number of earthquakes $N$ above a certain magnitude $M$. The b value, gives an estimate for the occurrence of large earthquakes compared to smaller earthquakes. Furthermore, its spatial distribution can be interpreted as a proxy for the distribution of seismic stress. Therefore, the b value has long been considered suitable to be used for earthquake forecasting, e.g.\ b value time series anomalies are used as precursors to large events \parencite{Smith1981, Main1989}, for earthquake rate forecasting \parencite{Smyth2011}, or to discriminate foreshocks from mainshocks \parencite{Gulia2019}.

Over the past decades, science has witnessed increased use of Deep Learning, especially leveraging the fast development in image processing \parencite{LeCun2015}. With the ever increasing amount of seismic data over time, this shift is also well-received in seismology and is successfully applied in tasks such as event detection and magnitude estimation using seismogram data \parencite{Chakraborty2022} or GPS (HR-GNSS) data \parencite{Quinteros2023}, seismic phase picking \parencite{Li2022}, synthetic waveform generation \parencite{lehmann2023} and more applications are likely to follow \parencite{Mousavi2022}. Deep Learning for Earthquake forecasting has equally produced some promising results recently \parencite{Shan2022, Herrera2022, Stockman2023, Fox2022}.

One of the most common methods in earthquake rate forecasting is the Epidemic Type Aftershock Sequence (ETAS) modelling \parencite{Ogata1993}. It is based on a point process to model the temporal activity, where a base rate of earthquakes generates aftershocks. While this model cannot be used to forecast main shocks, it is quite successful at estimating the rate of aftershocks.

However, there are no studies which combine the power of advanced deep learning architectures to handle larger data series and spatio-temporal b value distributions in order to do earthquake forecasting. In this work, we applied neural network architectures such as an autoencoder, Temporal Convolution and 2D convolutions to test the potential of deep learning in earthquake forecasting.

We use Japan as the study region which is visited by a high number of earthquakes every year, providing an ample amount of data to work with \parencite{Wakita2013}. Furthermore, it has an extensive network of seismic stations resulting in a good catalog completeness \parencite{Nanjo2010}. 
The tectonics of Japan is primarily characterized by the eastward subduction of the Pacific Plate beneath the Okhotsk Plate to the north and the Philippine Plate to the south. Concurrently, the Philippine Plate also undergoes subduction beneath the Eurasian Plate\parencite{Bird2003}. For the purpose of this study, we focus on the subduction zone of the Pacific and Okhotsk Plate.


The problem posed by a forecast of this type is twofold: (i) the limited training data, and (ii) the change of the underlying system by the seismic progression of the region. To account for this, we develop a progressive training routine using temporal increments to train the model continuously with a constant learning rate. With this approach, we do not have dedicated testing and validation sets, but rather a new testing-and-validation set for every new period. As we only train once, sequentially forward in time, no future information seeps into the model. 

\newpage
\section{Materials and Methods}
\subsection{Catalog Preparation}
\begin{figure}[H]
    \includegraphics[width=\textwidth]{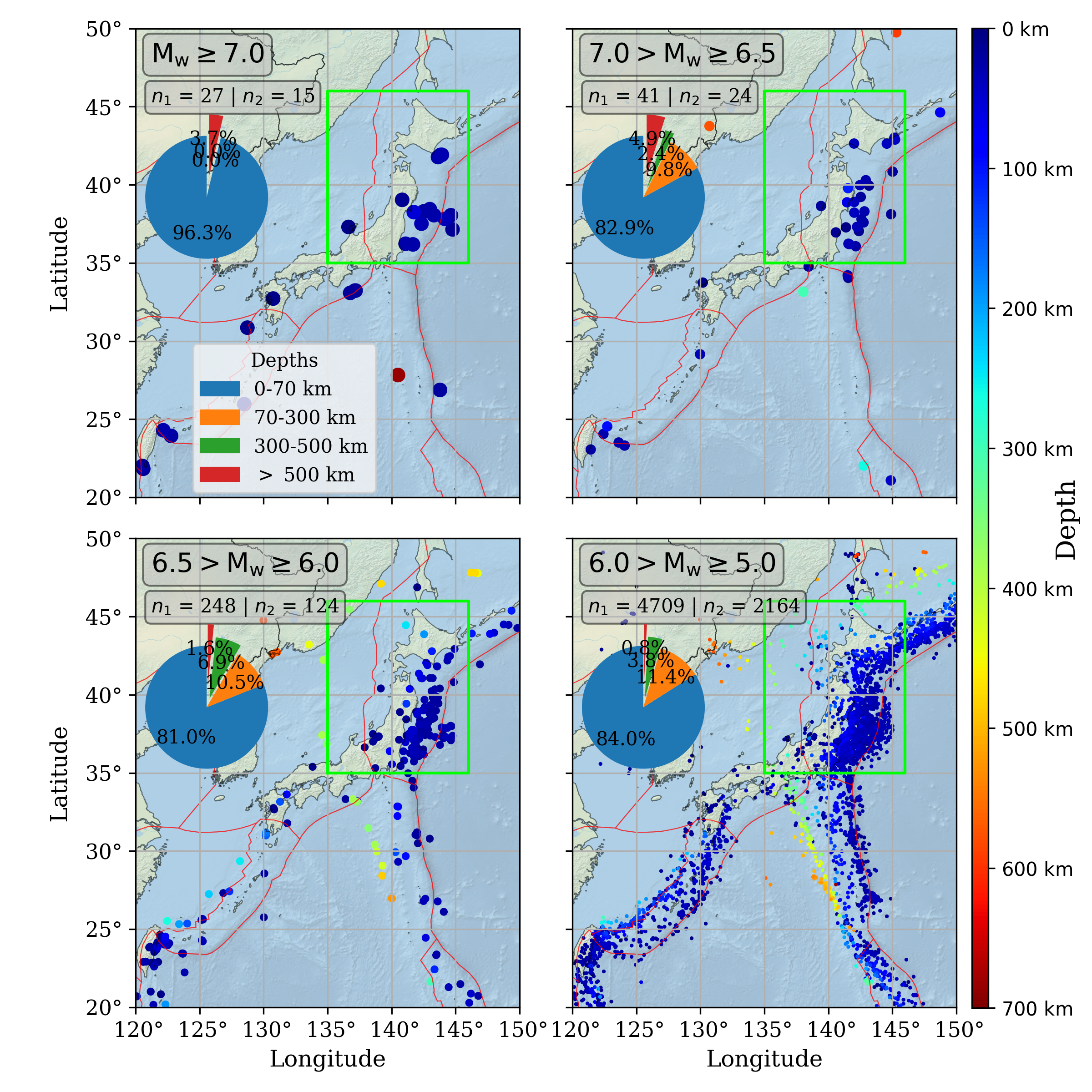}
    \caption{Distribution of earthquakes with Magnitudes \Mw $\geq 5$ from the beginning of our training data in 2001-05-27 to 2019-12-31.  The subplots show different magnitude bins. The inset pie chart shows the distribution of the depths, illustrating that the vast majority of stronger earthquakes happen in the shallow regions. The numbers $n_1$ and $n_2$ correspond to the number of earthquakes in the subplot or in the marked subregion of each plot. The red lines denoting the plate boundaries are taken from \parencite{Bird2003}.}
    \label{fig:Japan_Overview}
\end{figure}

For this study we focus our attention on the region $[20^\circ, 50^\circ] \times [120^\circ, 150^\circ]$ around Japan, as well as its subset $[35^\circ, 46^\circ] \times [135^\circ, 146^\circ]$ (See Figure \ref{fig:Japan_Overview}). This subset is chosen to simplify the system: it reduces the relevant plate boundaries from three to one, and limits the data to an area where most larger earthquakes occur in the shallow first 70~km.

We use the ISC Catalog \parencite{ISC_1of3, ISC_2of3, ISC_3of3} from 1999-01-01 to 2019-12-31 to create our dataset (accessed date is March 11th, 2022). ISC has better completeness than USGS catalog for our area of interest.

For our date and depth range, the catalog consists 3,111,016 events, recorded in a multitude of different magnitude types. We use the most common ones (\MS, \mb, \MV, \ML, \MJMA ) and convert them to moment magnitude \Mw, which is not widely present in the catalog, using the  relations given in \parencite{Sawires2019}. If the magnitude is given in more than one of the listed magnitude types, we prioritize the magnitudes in the order mentioned in the the bracket earlier. By restricting the use to only those magnitude types, we only lose 1884 events of smaller magnitudes ($\leq 3.5$) for the whole region.\\
Furthermore, we also limit the depth of earthquakes to 70~km, so only shallower earthquakes are considered. This has two main reasons: on the one hand, catalog completeness changes with depth and we deem it preferable to have a consistent catalog completeness, on the other hand, the b value empirically changes with the depth, which we do not take into account. Limiting the depth therefore should reduce the error we acquire by ignoring the depth in our b value calculation: We keep deep strong earthquakes from influencing the surface b value.
The dataset contains the 2011 Tōhoku Earthquake, a megathrust event which significantly changed the distribution of earthquakes in the area of interest. This is likely detrimental to the results but unavoidable, since the catalog before 1999 is less complete and the available time after 2011 is too short and still filled with an increased seismicity due to the aftershocks of the Tōhoku Earthquake.

\begin{figure}[H]
    \includegraphics[scale=1]{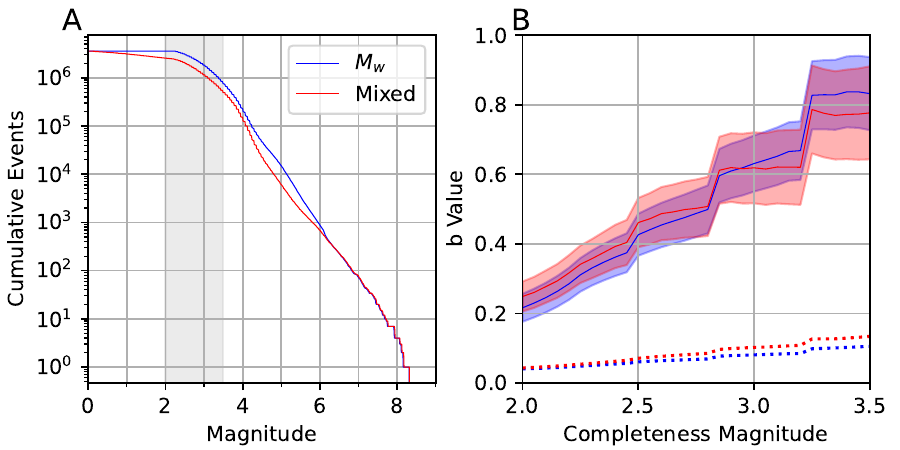}
    \caption{Magnitude conversion results. In A the cumulative number of earthquakes above a threshold magnitude are shown: The unconverted magnitudes of the ISC catalog in red and the converted magnitudes in blue. The gray shaded area contains the options for the completeness magnitude, which are further analyzed in B: The b value is shown there as a fit to $N=10^{a-bM}$) with the standard deviation of b in the shaded ares. The converted magnitudes have a smaller standard deviation, making them the better choice for further analysis. While the rest of this publication uses  Aki's method \parencite{Aki1965}, here the b value is shown as an exponential fit, because this will make the error more pronounced as it is less dependent on the completeness magnitude. The dashed lines show the absolute standard deviation.}
    \label{fig:MagnitudeConversion}
\end{figure}

The cumulative magnitude plot for converted and unconverted data is shown in Figure \ref{fig:MagnitudeConversion}, which shows a clear improvement in the linear behavior postulated by the Gutenberg Richter law. This is an important property, as we depend on the stability of the b value for our further analysis.

\subsection{b-Value Calculation}
Using the converted \Mw catalog we now calculate the b value on a fine $0.1^\circ$ by $0.1^\circ$ degree grid for each day after 2000-01-01. For each of those boxes, the b value is calculated using the relation from \parencite{Aki1965} on the earthquakes which occurred within a $0.25^\circ$ radius of the location center and within the last 365 days, akin to a cylindrical stencil. This yields a b-value array of size $7305 \times 300 \times 300$. If there are fewer than 2 earthquakes, we set the b value to 0, as well as in the case where all magnitudes for one cylinder have the same value. We clip the b value between 0 and 2, as there are some instances with three earthquakes and a very high b values. This method leads to unrealistic b-value results in the border regions of the seismic station coverage, as b=0 and b=2 cells border frequently, but empirically works well in this study.

\begin{figure}[H]
    \includegraphics[scale=1.3]{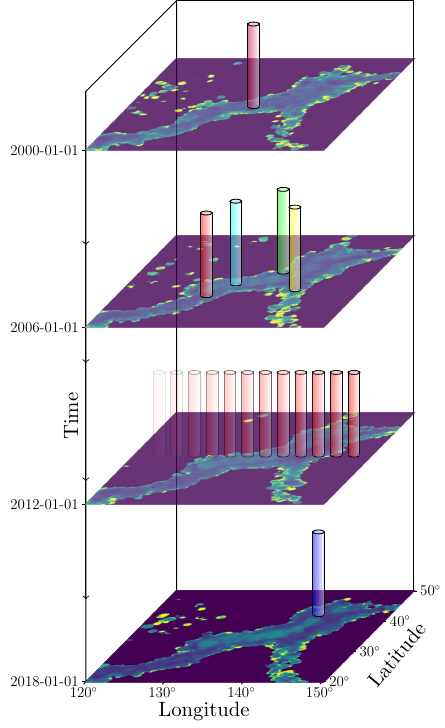}
    \caption{Workflow of the b-value calculation. For any given time and date a cylinder with $0.25^\circ$ radius going back one year in time is used to find the earthquakes from which this b value will be calculated. The different colored cylinders show different times and places. On the 2012 plane we try to visualize the cylinder rasterizing across the all the 3D distribution of magnitudes.}
    \label{fig:b_calculation_cylinders}
\end{figure}

\paragraph{Dataset Creation}
The creation of the dataset is represents an important step for a machine learning approach, because it embeds the information and format that the network should learn.
For our classification problem, we define two classes, each referring to a timseries of spatial b value reconstructions:

\vspace{-0.8em}
\begin{itemize}
\setlength{\itemsep}{-1.5em}
    \item[(1)] an earthquake class (referred to as \EQ), defined by an earthquake of magnitude \Mw $\geq 5$ on the day after the series ends\\
    \item[(2)] an \nEQ class, defined by the absence of such an event:
    \vspace{-0.8em}
    \begin{itemize}
        \setlength{\itemsep}{-0.2em}
        \item No earthquake with \Mw $ \geq 4.2$ within a $0.8^\circ$ L1 radius
        \item No earthquake with \Mw $ \geq 4.2$ within $\pm 7$ days
    \end{itemize}
\end{itemize}
The stronger conditions on the \nEQ class (as compared to ``anything that is not \EQ'') was introduced to increase the contrast between the two classes. The classification is performed on our own defined Deep Learning architecture which we test against other published networks.

As our b-value block starts on 2000-01-01 and we require 512 days of history for our classification later, the classification is limited to events after 2001-05-27 (2000-01-01 + 512 days) and 2019-12-31, in which time there are 4172 \Mw $\geq 5$ events, 2328 of which fall in the area used for training (green box, Figure \ref{fig:Japan_Overview}). For each of those events we carve a $512 \times 32 \times 32$ shaped sample centered around the epicenter of the event out of the b value block, so that the earthquake would happen on the 513th day, exactly one day after the sample block ends. The hypothesis here is that there are characteristic patterns in the b value prior to a large event \parencite{Main1989, Gulia2019}, and the network is supposed to find them. 
On the other side of the 4172 larger events, we also need a class that constitutes ``calm'' situations in which spatiotemporal sequences are not followed by a larger event.


For non-earthquake class we calculate all position for which all the following criteria (later referred to as SC) are met:
\begin{itemize}
    \setlength{\itemsep}{-0.4em}
    \item Within one week before and after the selected point as well as within $[-0.8^\circ, 0.8^\circ] \times [-0.8^\circ, 0.8^\circ]$ around the location there are no earthquakes with \Mw $\geq 4.5$.
    \item The b values in the $512 \times 32 \times 32$ block to be considered is calculated using at least on average 10 earthquakes.
\end{itemize}
From these possible locations we choose a number equal to the number of \EQ class events to maintain a balanced dataset for each training instance.

\begin{figure}[H]
    \includegraphics{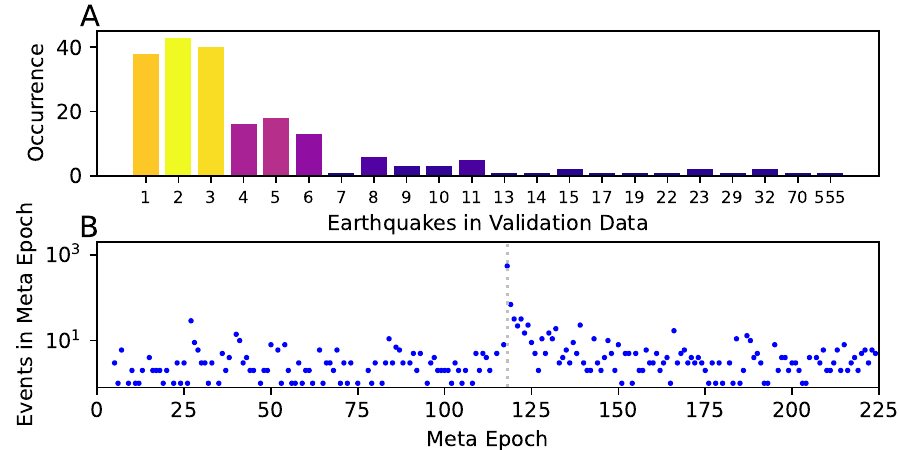}
    \caption{Earthquake Distribution in Validation Data. How often a number of earthquakes appears in the validation data is shown in A. Most meta epochs (explained in section \ref{sec:training}) contain 1 to 3 earthquakes, 4 to 6 are still common, and more earthquakes are rare. The very high numbers correspond to the Tōhoku earthquakes and the months that follow. This can be seen in B, where the number of events in each meta epoch is shown with over the meta epochs.}
    \label{fig:Res_Val_EQoccurrence}
\end{figure}


\subsection{Deep Learning Approach}
In principle we designed a two step approach similar to \parencite{Fox2021} for the classification of the previously defined b value blocks.
The idea is, that the autoencoder learns the ``normal'' state of the system, so when the system is in an abnormal state, the reconstruction error will be different. A second network, will then be used on the reconstruction error to classify the input between ``normal'' (\nEQ) and ``abnormal'' (\EQ) states.
Empirically we have found, that using this two stage architecture works better than using the classifier immediately on the raw data.

\begin{figure}[H]
    \includegraphics[scale=1]{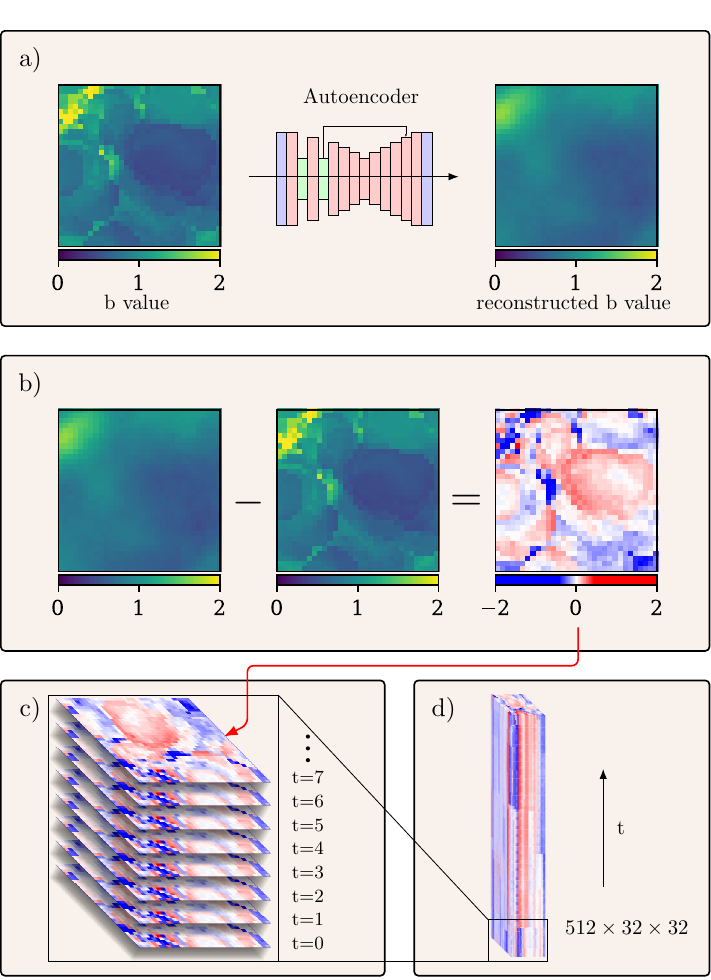}
    \caption{Pipeline from the b value grid to the input of the classifier network. First (a) all relevant $32 \times 32$ maps of the b value are sent through the autoencoder. In the second step (b) we take the difference between reconstruction and input to get a difference map. In the third step (c), 512 different maps are assembled in a sequential manner to form the final input block shown in d.}
    \label{fig:BlockCreation}
\end{figure}

In the first step we take the $[32 \times 32]$ b value array and run it through an autoencoder.
We then take the difference between reconstruction and input to create an array of reconstruction error. This is done for all 512 instances per block to create the input for our second step. The process is also illustrated in Figure \ref{fig:BlockCreation}.

The second step consists of a Convoluted Dilational Network (CDN), an architecture presented here and developed specifically to deal with 2D spatial and 1D temporal data. The CDN is used to classify the b valued difference blocks into the two categories of ``followed by a large earthquake'' (class \EQ) and ``not followed by a large earthquake'' (class \nEQ).
Both networks as well as their training processes are described in detail in the following two sections.

\subsubsection{Autoencoder}
The autoencoder (AE) is trained on a dataset of 250,000 images of $32 \times 32$ pixel values of the b value from the whole region of which samples could be taken in for the classifier. This includes all 365+512 days before an earthquake at that location (which would then be centered in the $32 \times 32$ image) as well as all locations that can possibly be selected according to the SC criterion defined above. $80\%$ of those images are used as training data, $20\%$ are used for testing.

The training is performed using mean squared error as the loss, and the model is trained until the validation loss does not improve for 50 epochs. 
We tried out different architectures for this task and a network based on dense layers, dropout, and skipped connections outperformed the alternatives (different variations of those three components and 2D Convolution layers).

The final model is as follows: The input is flattened to 1024 dimensions, and each subsequent dense layers cuts that in half. The smallest layer is 32, after which we double the dimensions until we reach 1024, which we reshape to $32 \times 32$. The first two layers have a dropout of 0.5, the skipped connection is after the second dropout layer dropout layer (256 dimensions) to the second to last layer.

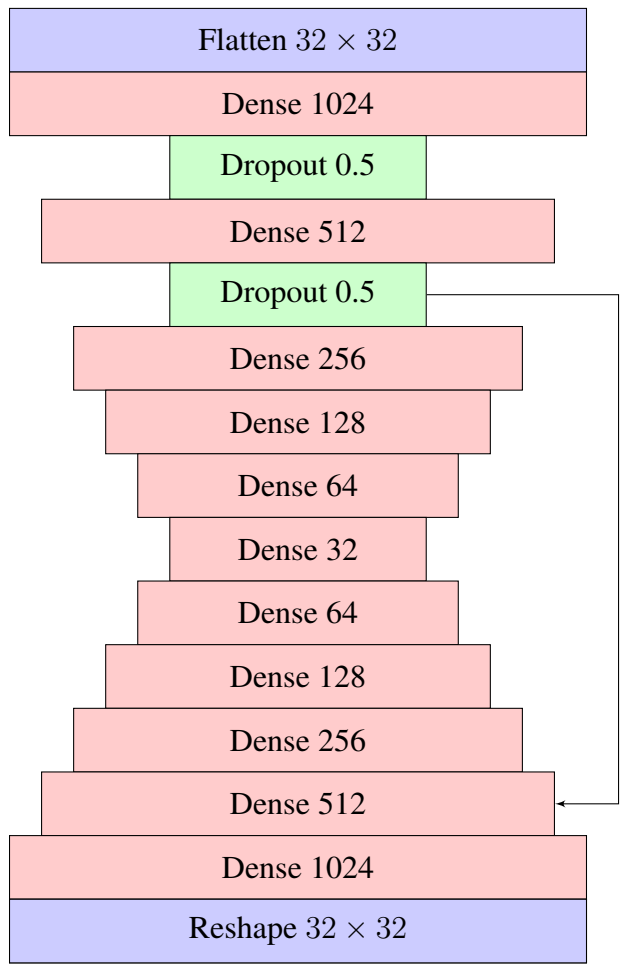
\begin{figure}[H]
    \tikzstyle{block} = [draw, fill=red!20, rectangle, minimum height=2em, minimum width=8em]
    \tikzstyle{coord}  = [coordinate, on grid]
     \begin{tikzpicture}[node distance=2em,>=latex', scale=0.4]
        \node [block, minimum width=18em, fill=blue!20]  (f1) {Flatten $32\times32$};
        \node [block, minimum width=18em, below of=f1]  (d1)  {Dense 1024};
        \node [block, minimum width=8em, below of=d1, fill=green!20]  (do1)  {Dropout 0.5};
        \node [block, minimum width=16em, below of=do1]  (d2)  {Dense 512};
        \node [block, minimum width=8em, below of=d2, fill=green!20]  (do2)  {Dropout 0.5};
        \node [block, minimum width=14em, below of=do2]  (d3)  {Dense 256};
        \node [block, minimum width=12em, below of=d3]  (d4)  {Dense 128};
        \node [block, minimum width=10em, below of=d4]  (d5)  {Dense 64};
        \node [block, minimum width=8em, below of=d5]  (d6)  {Dense 32};
        \node [block, minimum width=10em, below of=d6]  (d7)  {Dense 64};
        \node [block, minimum width=12em, below of=d7]  (d8)  {Dense 128};
        \node [block, minimum width=14em, below of=d8]  (d9)  {Dense 256};
        \node [block, minimum width=16em, below of=d9]  (d10) {Dense 512};
        \node [block, minimum width=18em, below of=d10] (d11) {Dense 1024};
        \node [block, minimum width=18em, below of=d11, fill=blue!20] (f2) {Reshape $32\times 32$};
        \node [block, minimum width=18em, below of=d11, fill=blue!20] (f2) {Reshape $32\times 32$};
        \node[coord, right= 10em of do2] (h1) {};
        \node[coord, right= 10em of d10] (h2) {};
        \draw [->] (do2) -- (h1) -- (h2) -- (d10);
    \end{tikzpicture}
    \caption{Autoencoder Architecture. This figure shows the autoencoder for the b value arrays, with skipped connections and dropout.}
    \label{fig:AE_Architecture}
\end{figure}

\subsubsection{Classifier}
The second part of our forecasting pipeline consists of a classifier that takes a timeseries of the reconstruction error of the autoencoder. The data should be classified in two categories, referred to as \EQ (earthquake with a magnitude $\geq$ 5 \Mw) and \nEQ (data selected according to SC).

This gives the network spatiotemporal information about the b value leading up to the point of interest, $1.6^\circ$ degrees in either direction and 512 + 365 days (with an overlap of 365 days) in the temporal axis.

\begin{figure}[H]
    \includegraphics[width=0.45\textwidth]{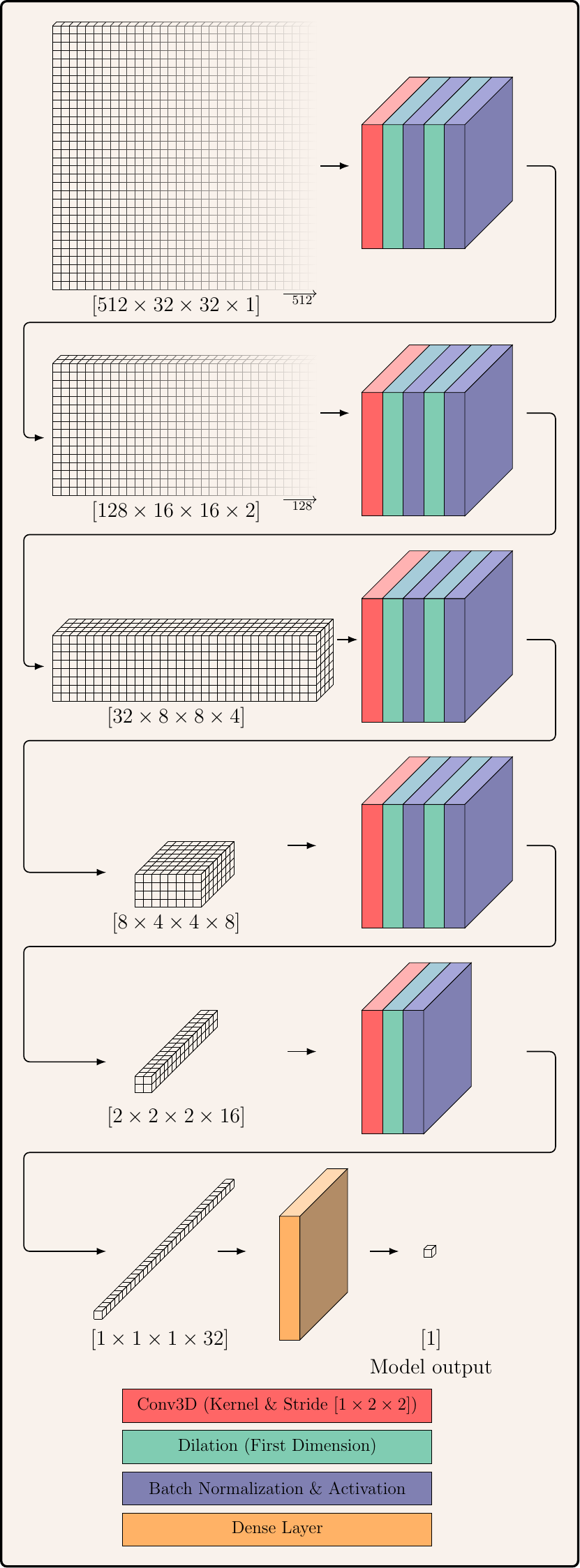}
    \caption{Architecture illustration for the Convolutional Dilated Network. The CDN is used to reduce the input data from  $512 \times 32 \times 32 \times 1$ dimensions to 1. Since the second and third value for the dimensionality of that tensor are always the same, we collapse them to one dimension in the sketch. The 3D convolutional layer transforms its input dimensionality from $i \times j \times j \times k$ to $t\times \frac{j}{2} \times \frac{j}{2} \times 2k$, while the dilation is a 1D convolutional layer with a dilation set to the powers of 2. This reduces the dimensionality in the first value by a factor of 2 each time it is called. The model output is related to the probability of a large event occurring.}
    \label{fig:CDN}
\end{figure}

The architecture is the combination of 1D convolutional layers in time (TCN \parencite{Lea2017}) and 2D convolutional layers in space. 
The spatial dimensions are reduced with 3D Convolutional layers (with a kernel of shape $[1 \times 2 \times 2]$ in order to only act on the spatial dimensions) and the temporal dimension is reduced using 1D Convolutional layer of increasing dilation. Between the different types of convolutions we apply the necessary reshaping operations, and every 1D convolutional layer is followed by an activation layer (Leaky ReLu with $\alpha = 0.1$). Due to the changing dimensions we did not use the skip connections as a ResNet\parencite{He2016} architecture  would, although this could be implemented by way of a maxpooling layer as an alternative path to the convolutions. The arcitecture, as well as the intermittent dimensionalities of the data are shown in Figure \ref{fig:CDN}. After the final convolution layers we have 32 channels, which are reduced to a single output using one dense layer.

\subsubsection{Architecture Comparison}
\begin{figure}[H]
    \includegraphics[width=1.0\textwidth]{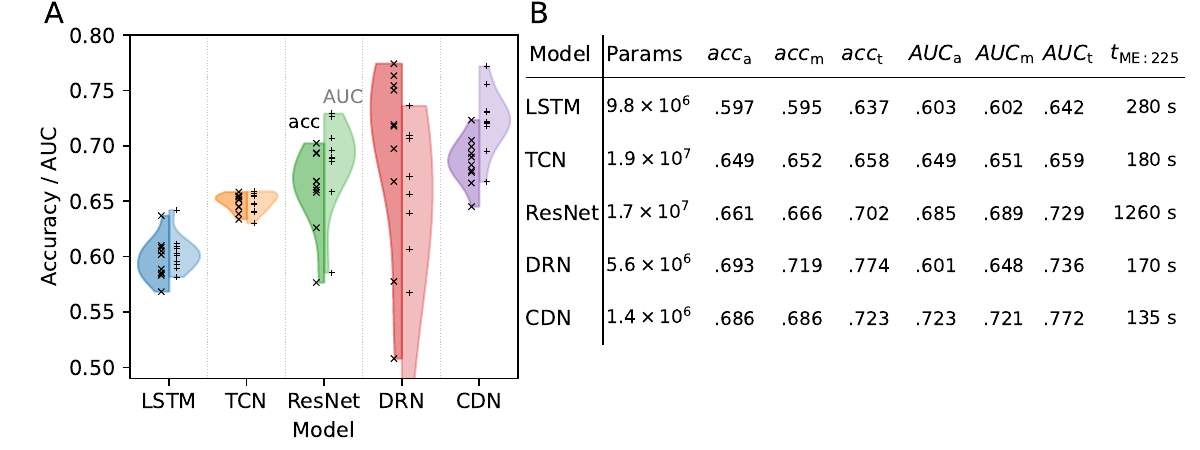}
    \caption{Architecture Comparison. Panel A shows the accuracy (acc, left) and area under the curve (AUC \parencite{Ling2003}, right ) for the different architectures. Every architecture was initialized with 10 different seeds. The individual results are shown as $\times$ for the accuracy and as $+$ for the AUC. Panel B shows the number of parameters for the models, as well as the mean (${acc}_\mathrm{a}$), median (${acc}_\mathrm{m}$) and top (${acc}_\mathrm{t}$) values for the accuracy and the AUC.}
    \label{fig:Architecture_Comparison}
\end{figure}

The CDN architecture used in this paper was tested against several state-of-art deep neural networks for similar data structures. An overview of the results, parameters and speed is shown in figure \ref{fig:Architecture_Comparison}.

\textbf{Temporal Convolutional Network (TCN)} \parencite{Lea2017}. This architecture consists of 9 1D Convolutional layers with increasing dilation, reducing the temporal dimension from 512 to 1. Initially, the input of a single day is flattened to a 1D array, which is retained throughout the Temporal convolution. As a last step, the 1024 dimensional array is reduced to one value.

\textbf{Long Short-Term Memory (LSTM)} \parencite{Hochreiter1997}. For the LSTM Network we use 4 Layers with 1024, 256, 64, and 16 units, the first three of which return sequences. The last layer is a dense layer, reducing the output to one dimension.

\textbf{ResNet}  \parencite{He2016}. Our 3D data with the temporal component is not an ideal input for ResNet, as the temporal component is not treated in an appropriate way. For this architecture we use $7 \times 3$ ResNet Blocks with the channels doubling every three blocks (starting with 1 channel). One block consists of twice, a 2D Convolutional layer, batch normalization, and a ReLU activation. After each group of three dense ResNet Blocks, we use a Maxpooling layer to reduce the temporal extent by a factor of two (this is not in accord with the original network from cite \parencite{He2016}, but necessary to reduce the dimension of the problem to accommodate hardware limitations). Each ResNet Block can be skipped. 

\textbf{Dilated Recurrent network (DRN)}. With the Dilated Recurrent Network we combine the approaches of ResNet and TCN.
For this we use 1D Convolutions in time (TCN-like) while also providing skipped connections via maxpooling in time.
The DRN blocks consists of twice, a 1D convolution layer, batch normalization, and activation, while the skipped connection via maxpooling layer connects just before the last activation. To achieve the correct dimensions the maxpooling has a pool size of 2 and a stride of two. This DRN is not to be confused with the Dilated Recurrent Neural Networks from \parencite{Chang2017DRN}, but there are some similarities.

\textbf{Convolutional Dilated Network (CDN)}. The Convolutional Dilated Network is an improvement on the Dilated Recurrent Network and the basis for the results presented in this paper. The goal while designing this network was to combine spatial reduction from the TCN while at the same time keeping the 2D spatial features of the data, instead of flattening the spatial dimensions as in the DRN.
Reshaping the data between 1D Convolution from the TCN and the 3D Convolutions for the feature extraction adapts the network to the given problem. Similar to TCN and DRN, the CDN uses 9 blocks to reduce the 512 dimensional time axis to 1. 
At the same time, to reduce complexity, we reduce the spatial dimensions every second block (starting wit the first one). A CDN block starts with a 3D Convolutional layer with a (1,2,2) kernel and a (1,2,2) stride for the odd numbered blocks, and continues for all blocks with a reshape to enable the 1D Convolution, a reshape back, a batch normalization and an activation.  The 3D convolution increases the number of channels by a factor of two, the 1D Convolution keeps it constant.
The final 32 dimensional output is reduced to a scalar with one dense layer.

Even though the DRN has the highest overall accuracy, we use the CDN in this paper, for the following reasons:
\begin{itemize}
    \setlength{\itemsep}{-0.4em}
    \item The results are more stable between runs.
    \item The training is faster.
    \item The network uses less parameters.
    \item The area under the curve (AUC) of the ROC is larger, which points to a more stable threshold and is generally regarded as a better measure for the evaluation of models \parencite{Ling2003, Huang2005}.
\end{itemize}

\subsubsection{Training the Classifier}
\label{sec:training}
\begin{figure}[H]
    \includegraphics[width=\textwidth]{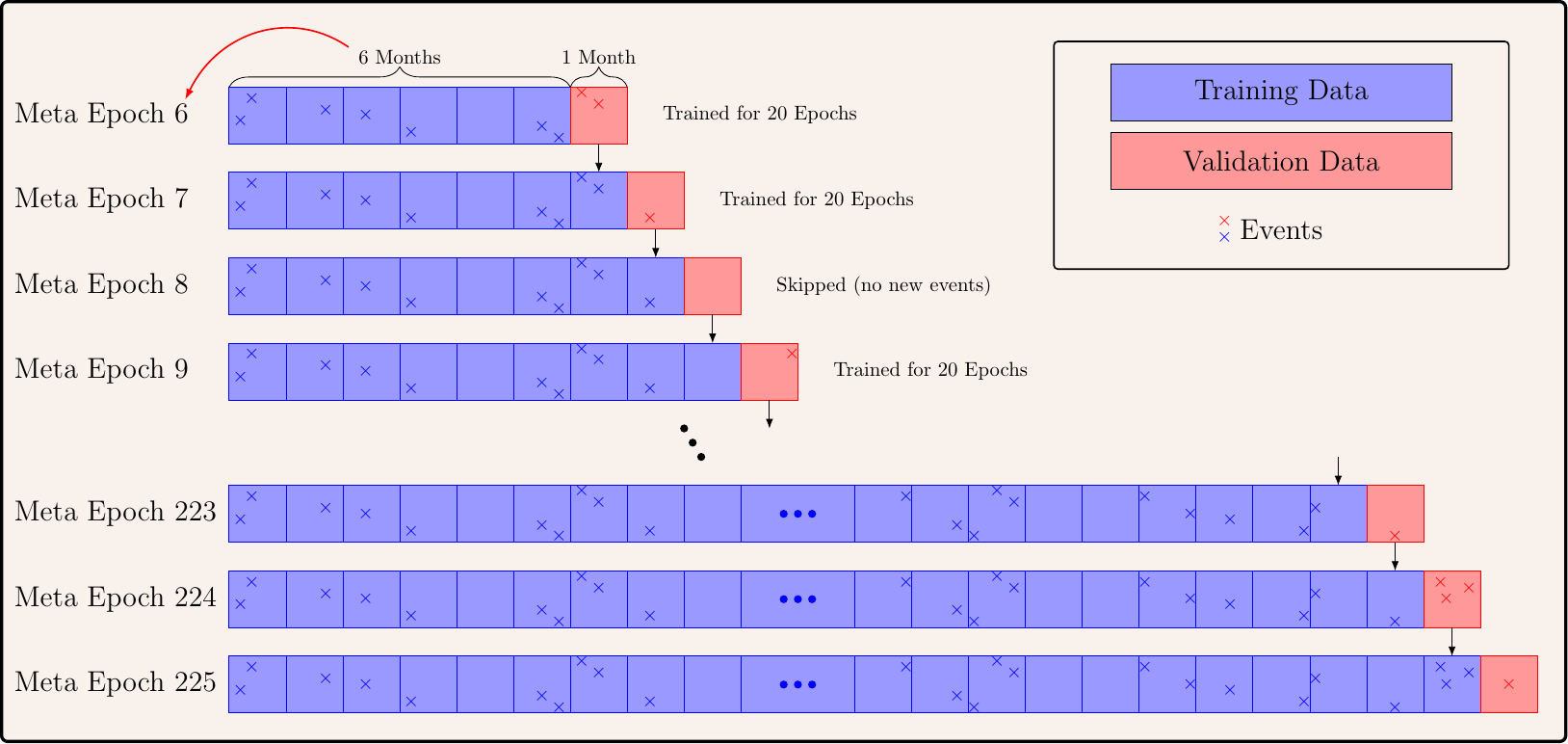}
    \caption{Training process illustration. This figure shows the changing datasets used for training and validation during the whole training process. The validation data of a previous meta epoch is added to the training data of the next meta epoch, if there are any large earthquakes in that data. Meta epochs without new \EQ class events are skipped.}
    \label{fig:TrainingSketch}
\end{figure}

Training the classifier without violating causality and in a way that can be used in applications required an evolving training and testing set, Figure \ref{fig:TrainingSketch} shows how this is accomplished. The training process is divided in \textbf{meta epochs}.
A meta epoch refers to both, a training cycle consisting of 20 epochs of normal training, and the temporal extent of the validation data, which is 30 days.

In order to progressively train the model on newer data, each meta epoch trains and validates on new data, which is also why we do not use a dedicated testing set in this work.
At the beginning the whole time domain is split in chunks of 30 days. 
Then training is performed on an increasing number of those 30 day chunks / meta epochs.
Training starts with  6 such chunks (because starting with less means starting with an almost empty training set).
Therefore the first meta epoch is labeled meta epoch 6, because it contains 6 chunks.
Each meta epoch is trained for 20 normal epochs. 
The training and validation sets change each meta epoch (See Figure \ref{fig:TrainingSketch}):
The training data contains all \EQ class events up to the current time, as well as a random selection of allowed \nEQ class events from the possible locations allowed by the SC criteria defined above. This exposes the network to a wide variety of \nEQ events over time.
In order to prevent overfitting for older events, to which the network would be exposed more often compared to newer events, we weigh all samples from the training data that were not newly added this meta epoch using a weight so that the total new data equals (in weight) the old data.

The validation set is the chosen from the next meta epoch. It contains all \EQ events of that 30 day period as well as an equal amount of randomly chosen \nEQ events from the same period.

\newpage

\section{Results}

The training and validation here are done in 30 day intervals, where the model is trained on some amount of data, validated on the next 30 days, and then trained on the combined data, and trained on the next 30 days. This training method is explained in more detail in Figure \ref{fig:TrainingSketch}. Owing to this specialized training procedure, giving a normal model evaluation such as the final accuracy is not as informative compared to other models. We therefore name the single training periods \textbf{meta epochs} and report the relevant results for those meta epochs instead of a total result.

\begin{figure}[H]
    \includegraphics[width=\textwidth]{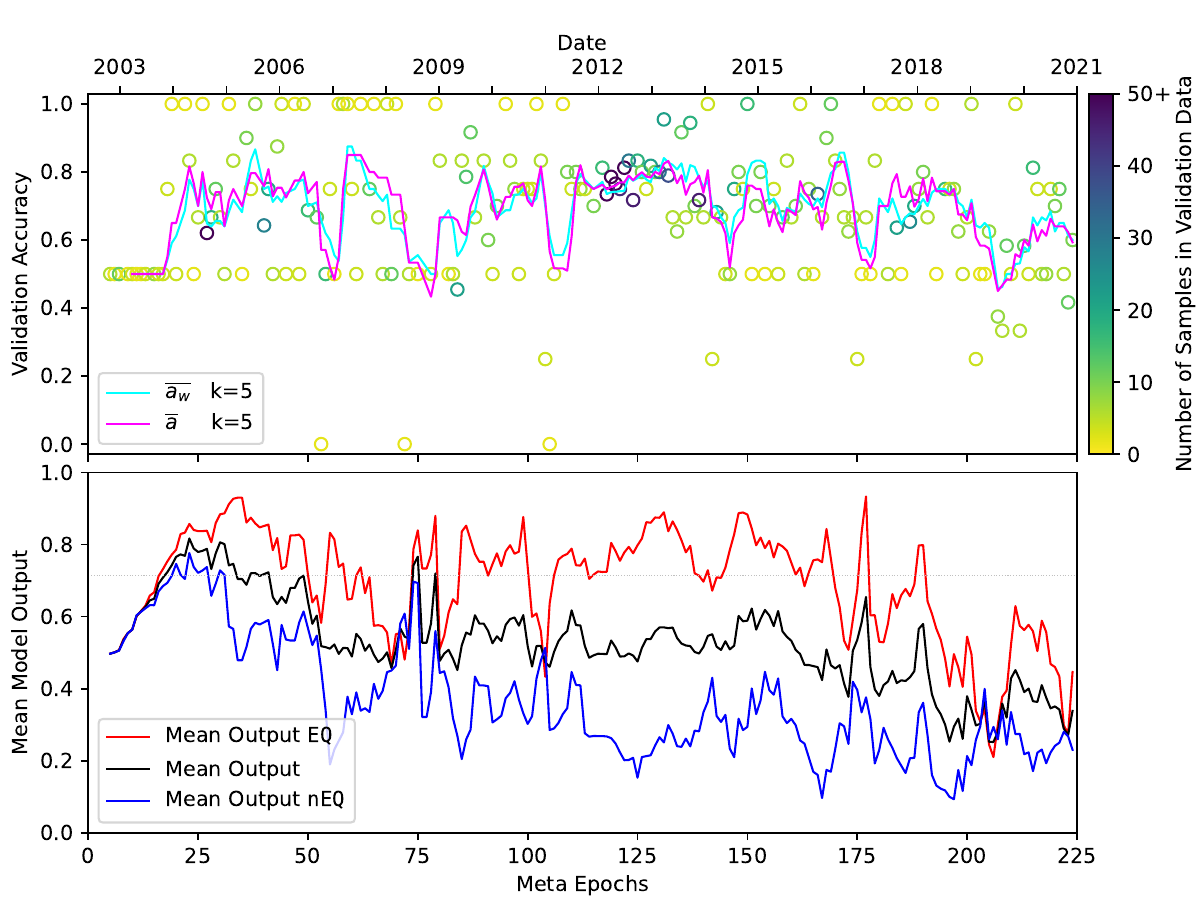}
    \caption{Classification Accuracy for meta epochs. The markers each correspond to one meta epoch and are colored based on the number of samples in the validation set for that meta epoch, however we limited the color range from 0 to 50. There are single meta epochs with much higher counts, especially those following to the Tōhoku Earthquake. A closer look at the Events in the validation data is provided in Figure \ref{fig:Res_Val_EQoccurrence}. The lines show different running averages for the accuracy: The light blue is a weighted average (since the meta epochs contain different numbers of samples), the magenta lines are a non weighted average. The average is taken with a kernel of 5 meta epochs.
    The lower panels shows the model output (smoothed over 5 meta epochs and weighted) for the two classes and the validation set as a whole. Generally, the output for the \EQ class is higher, as expected, however there are some instances where the three lines meet. The grey horizontal line corresponds to the best threshold (see \ref{fig:Res_Acc_CM_ROC}) for discrimination between the classes.}
    \label{fig:Res_Acc_ME}
\end{figure}

As the progressive training changes the model and the model is applied to new data for each determination of the accuracy, the overall accuracy of the model is difficult to determine. However, the accuracy is the most common metric to measure the model performance, so Figure \ref{fig:Res_Acc_ME} shows the accuracy over time. The time intervals are given in increments of 30 days, which also correspond to the progressive training cycles and are referred to as meta epoch from now on. As the 30 day forecasting period (= 1 meta epoch) usually contains only one to three earthquakes (see Figure \ref{fig:Res_Val_EQoccurrence} for more details on the distribution), resulting in rather quantized accuracy values, Figure \ref{fig:Res_Acc_ME} shows the accuracy averaged over 5 meta epochs as well.  Figure \ref{fig:Res_Acc_ME} shows, that after a few meta epochs at the beginning the accuracy improves to $75\%-80\%$ and only declines during the last two years to around $60\%$.

\begin{figure}
    \includegraphics[scale=1]{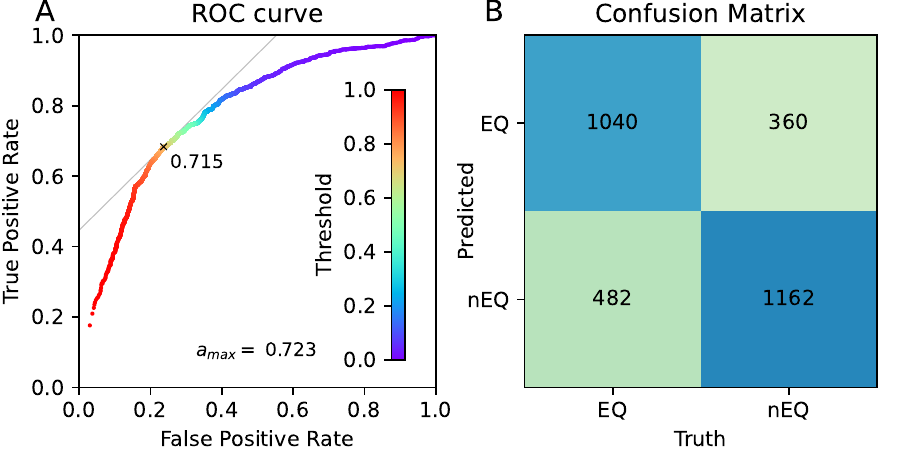}
    \caption{ROC Curve and Confusion Matrix. This figure shows the ROC Curve (A) and the confusion matrix (B). Using a classification threshold of 0.715, the model yields the highest accuracy: $72.3\%$. The confusion matrix shows a slight bias towards the \nEQ class, but an overall significant performance.}
    \label{fig:Res_Acc_CM_ROC}
\end{figure}

The optimal discriminating threshold between the \EQ and \nEQ classes is not necessarily 0.5 (with a model output between 0 and 1), and the Receiver Operator Characteristic (ROC) \parencite{Bradley1997} helps to determine the best value to classify the input by plotting the true positive rate against the false positive rate, allowing to find the optimal combination of both. Figure \ref{fig:Res_Acc_CM_ROC} shows the ROC as well as the resulting best confusion matrix \parencite{Ting2017}, which displays the number of true/false positives/negatives. The overall accuracy of the model, including the caveats discussed above, is $72.3\%$, including the initial ``untrained'' phase. It is calculated on the final model of each meta epoch, with the validation data of that epoch.

\begin{figure}[H]
    \includegraphics[scale=0.8]{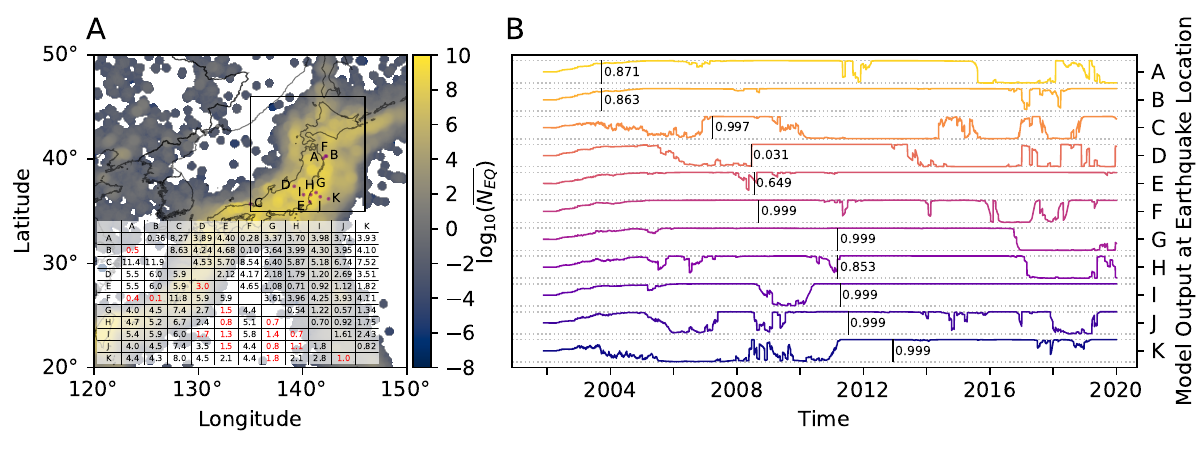}
    \caption{Location and Model output for earthquakes $\geq 7.0$. Panel A shows the average number of earthquakes used to calculate any one b value at that location and the respective L1 (lower left) and L2 (upper right) norm distances between the major earthquakes. Red marked distances imply an overlap in the input data for those earthquakes. Panel B shows the model output at the location of all earthquakes as well as the time of the major earthquake at that location and the model output for that day.}
    \label{fig:Res_EXE_Local_Prediction}
\end{figure}

\begin{figure}[H]
    \includegraphics[scale=0.8]{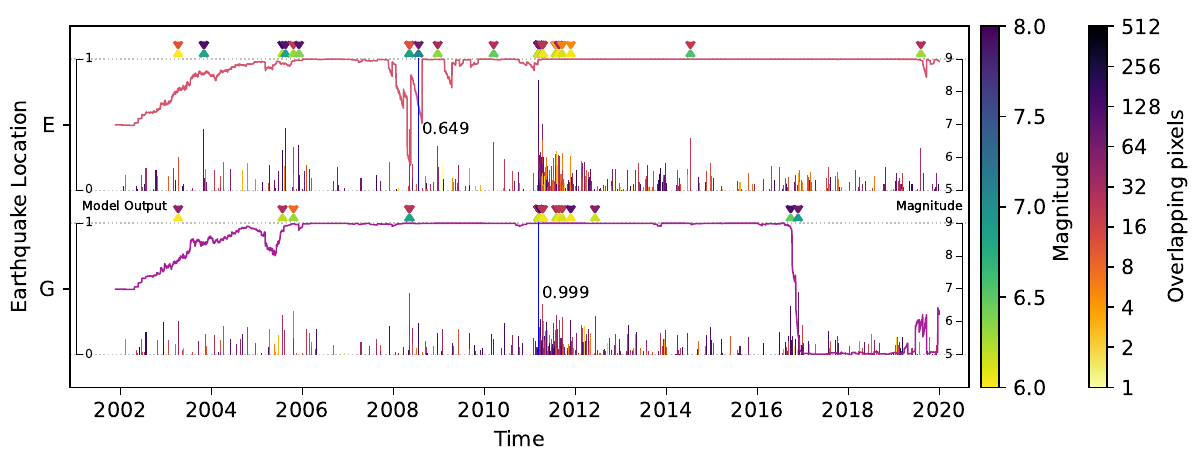}
    \caption{Zoom-in visualization of  the model output on Earthquakes E and G from Figure \ref{fig:Res_EXE_Local_Prediction}. The same figure for all earthquakes is in the supplemental material. The contiguous line corresponds to the model output (left scale). The vertical line with the numbers behind it denotes the timing of the large earthquake at that location. Smaller vertical lines signify earthquakes close to the the main earthquake, the height of the line corresponds to their magnitude (right scale), the color corresponds to the distance (right colorbar). The markings above the model output area illustrate again the earthquakes with \Mw $\geq 6$, the color of the lower triangle corresponds to the magnitude (left colorbar), the upper triangle's color to the overlap (right colorbar).}
    \label{fig:Res_Acc_PositiveAnalysis}
\end{figure}

\textbf{Targeting the \Mw $\geq$ 7 Earthquakes}\\
Since forecasting earthquakes with \Mw $\geq 7.0$ correctly is a priority from a hazard analysis point of view, we also target the 11 earthquakes with \Mw $\geq 7.0$, listed in Table \ref{tab:EQs_ge7}, in addition to the \EQ class. 
Those earthquake locations are shown in panel A of Figure \ref{fig:Res_EXE_Local_Prediction}. In order to see how the model output changes over time for those locations, we also show the model output for the whole time domain at the earthquake's location, with the model parameters being set to correspond to the time of the forecast, so that no future information is used. This is seen in the panel B of figure \ref{fig:Res_EXE_Local_Prediction}, together with the time of the earthquake (vertical line) and the model output for that day.
This illustrates the abilities and drawbacks of the model: Generally the output is quite high in high seismicity regions. In the beginning, the untrained model learns fast to increase the output , which can be seen in all traces from the beginning of 2002 to the end end of 2003. However, then the model adapts and the forecasts begin to differ. Earthquake C is captured quite well, the model output increases just before the earthquake occurs and decreases afterwards. The other earthquakes need closer examination, which is provided in Figure \ref{fig:Res_Acc_PositiveAnalysis}. Here, the model output for earthquakes E and G is shown in greater detail. The model output is augmented with the distance and magnitude of nearby earthquakes: the right scale shows the magnitude of nearby earthquakes, which are shown (if \Mw $> 5$) as vertical lines in the respective model output, their height signifying their magnitude on the right scale. The color corresponds to the overlap in pixels for the $[32 \times 32]$ geographical location. Earthquakes with \Mw $> 6$ are also marked again above the model output for greater color perception by the reader, with the color of the upper triangle corresponding again to the overlap between the presented earthquake and nearby earthquakes, and the lower triangle corresponding to the magnitude. The respective colormaps are on the left and consistent for the figure.
A complete version for all 11 Earthquakes can be found in the supplemental material, see Figure \ref{fig:Res_Acc_PositiveAnalysis_all}.

\begin{table}[H]
    \centering
    \begin{tabular}{l|l|c|c|l}
    Day & Date & \Mw (USGS) & \Mw (Converted) & Name or (Region) \\
    \hline
    1364  &  2003-09-25  & 8.3 & 8.10  & 2003 Tokachi Earthquake \\
    2641  &  2007-03-25  & 6.7 & 7.04 & 2007 Noto Earthquake \\
    3087  &  2008-06-13  & 6.9 & 7.03 & 2008 Iwate–Miyagi Nairiku earthquake \\
    3123  &  2008-07-19  & 7.0 & 7.11 & (E of Namie)\\
    3177  &  2008-09-11  & 6.8 & 7.02 & (SSE of Obihir) \\
    4086  &  2011-03-09  & 7.2 & 7.49 & 2011 Tōhoku earthquake precursor \\
    4088  &  2011-03-11  & 9.1 & 8.35 & 2011 Tōhoku earthquake  \\
    4115  &  2011-04-07  & 7.1 & 7.02 & April 2011 Miyagi earthquake \\
    4209  &  2011-07-10  & 7.0 & 7.10  & (ESE of Ishinomaki) \\
    4725  &  2012-12-07  & 7.3 & 7.30  & 2012 Sanriku earthquake \\
    5047  &  2013-10-25  & 7.1 & 7.09 & (East Coast of Honshu) \\
    \end{tabular}
    \caption{This Table shows the 11 earthquakes with \Mw $\geq 7.0$, their day after the start of out data, the date, the USGS Magnitude \Mw, our converted \Mw Magnitude and the Name or region of the earthquake. While there is some disagreement in the magnitudes due to the conversion using \parencite{Sawires2019}, this is not expected to noticeably influence the results.}
    \label{tab:EQs_ge7}
\end{table}

\begin{figure}[H]
    \includegraphics[scale=0.6]{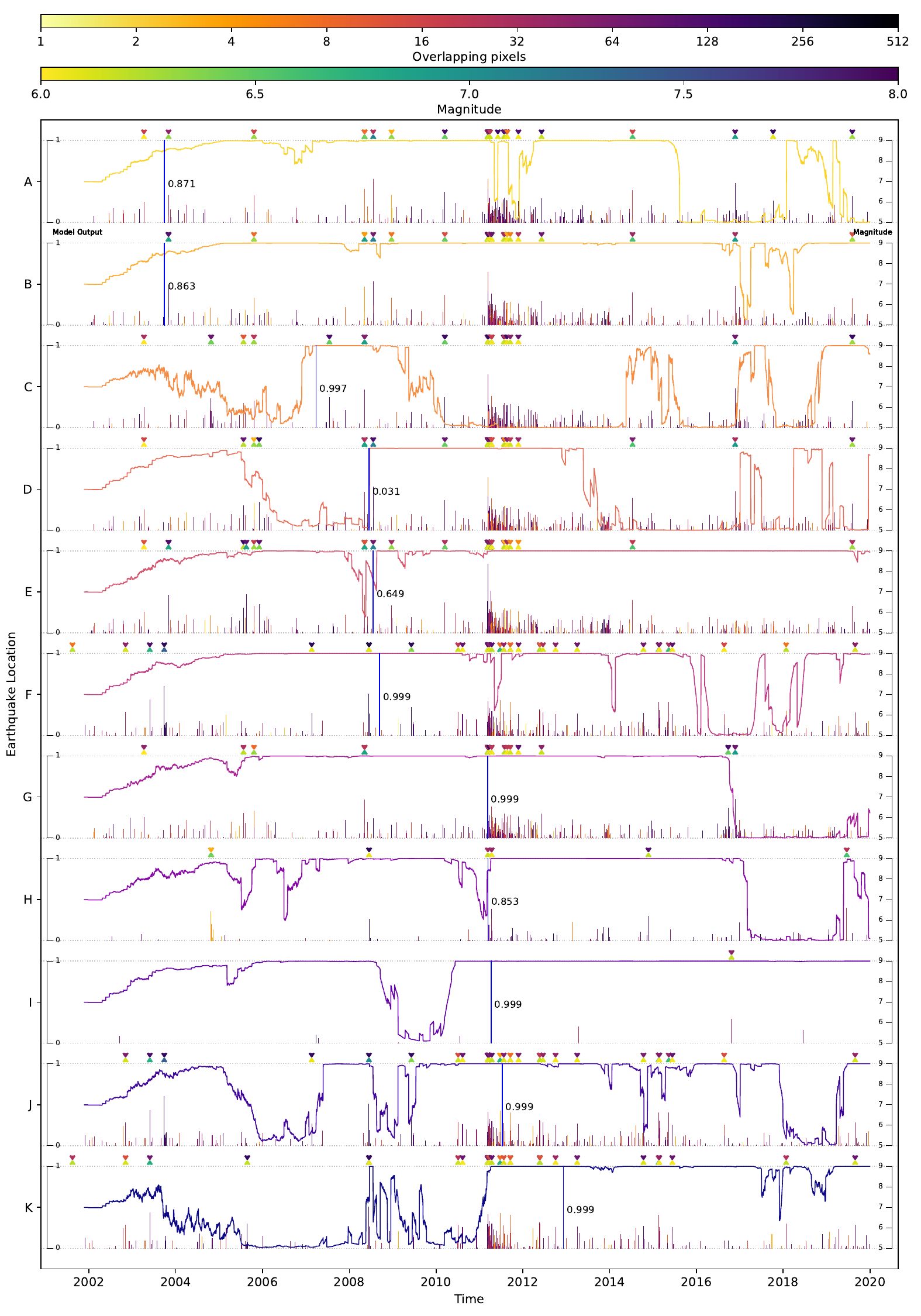}
    \caption{Position analysis for all $M \geq 7$ Earthquakes. See Figure \ref{fig:Res_Acc_PositiveAnalysis} for explanation.}
    \label{fig:Res_Acc_PositiveAnalysis_all}
\end{figure}

\section{Discussion}
In order to investigate if the overall forecasting ability of the model is valid rather than an assortment of artifacts we will have a closer look at several possible sources for errors: magnitude, depth, quality (or the number of earthquakes used for the b Value calculation), and location.

\subsection{Magnitude Dependence}\ 
If the premise, that the probability of an earthquake happening can be forecast by using changes in b value, it would suggest, that larger events are easier to forecast, simply because there are innumerate small earthquakes with \Mw $< 3$ that do not all have some large pattern of b value changes as a precursor. 
However, the model is trained mostly on forecasting smaller earthquakes, as the validation data contains mostly those (following the Gutenberg-Richter law \parencite{Gutenberg1944}), so it is a priori unclear, if the model can work for large earthquakes.
In order to test this, we analyze the magnitude dependence of the networks model outputs, as shown in Figure \ref{fig:Res_Acc_triple} (A). The left plot shows the prediction value for all \EQ samples, their magnitude and their meta epoch. While they cluster around 0 and 1, there is a clear trend towards 1. The mean prediction of \EQ samples averages over five meta epochs is shown in Figure \ref{fig:Res_Acc_ME} B, clearly showing good results for until roughly meta epoch 180. 
The right plot shows the prediction distributions for six magnitude bins, the center line is marked with the actual predictions, and the mean (colored) and average (black) for each bin are marked. It seems like smaller earthquakes are actually easier to forecast, which is probably due to two reasons: (1) the pattern that the network recognizes depends on the size of the earthquake, which seems likely, but since there are fewer large earthquakes, the network does not learn their pattern as well, or (2) smaller earthquakes are more commonly aftershocks and the networks just learns to predict more earthquakes in a periods after larger earthquakes. While (2) would be detrimental to our efforts, not all the performance could be explained away using this argument, because even the larger events (which are not aftershocks) are generally well predicted with significant accuracy.

\subsection{Depth Dependence}\ 
Earthquakes vary with depth and location, in our region they appear at higher depths in the subducting pacific plate, as well the shallow zones of the Okhotsk plate in the west and the pacific plate in the east. Depth is not a parameter the model is given in any way (apart from the depth limit of 70 km in the b value calculation), so in principle mixing different origins of earthquakes could pose a problem. Figure \ref{fig:Res_Acc_triple} B shows the effect of the depth on the model output, which seems to be negligible, meaning there is no bias that only deep or shallow earthquakes are treated correctly.

\subsection{Quality Dependence}
\begin{figure}[H]
    \includegraphics[scale=0.78]{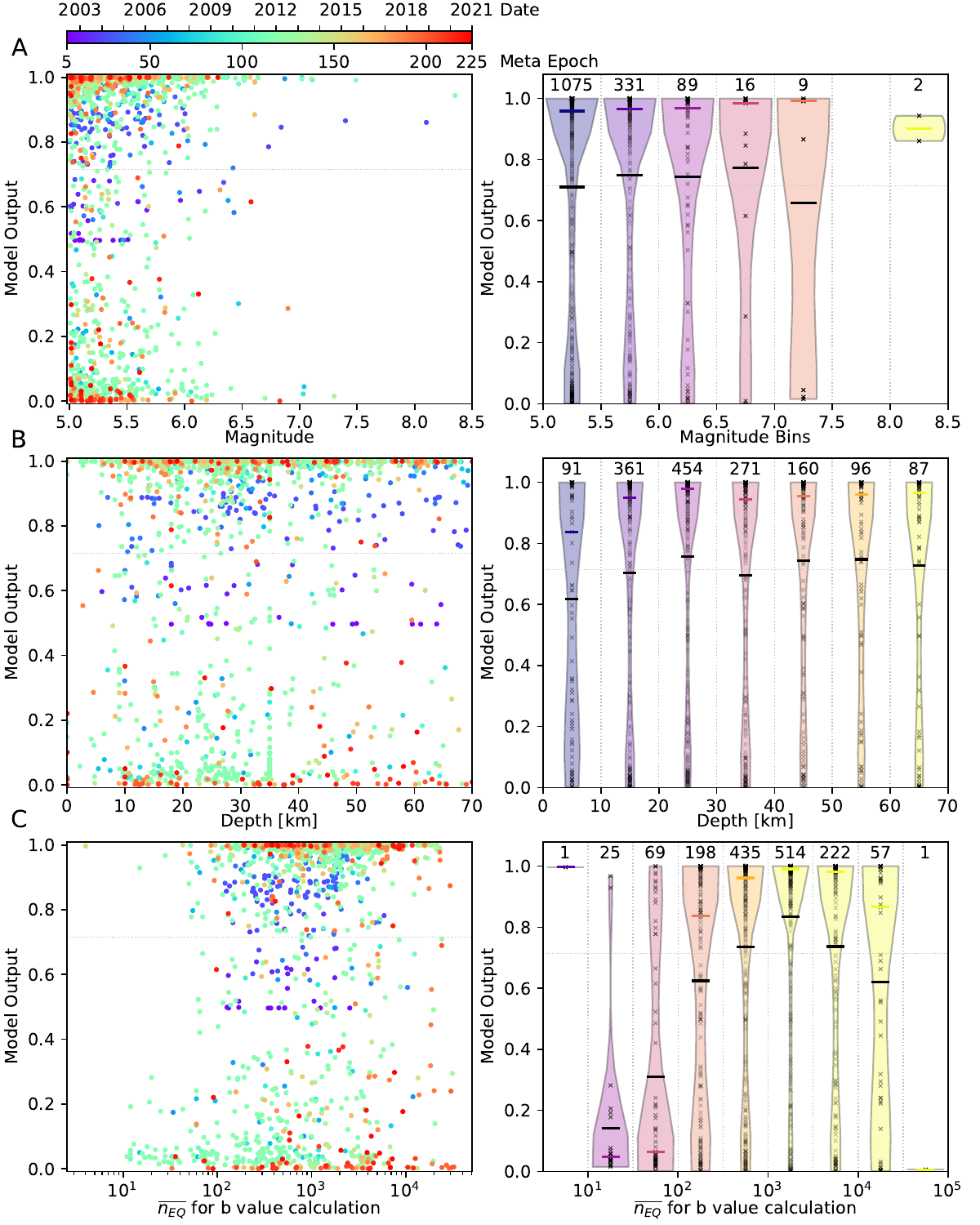}
    \caption{Impacts of magnitude (A), depth (B), and quality (C) on the model output. Each quantity's influence on the model output are shown, once over time and once in bins of that quantity. The gray horizontal lines corresponds to the optimal threshold from the ROC curve. In the left plots, the color corresponds to the meta epoch. In the binned plots, the colored bars correspond to the median, the black bars to the mean.}
    \label{fig:Res_Acc_triple}
\end{figure}

\ 
The quality of the b value data used for the forecasting process is bound to have an influence on the accuracy of the forecasting. Quality here is defined as the certainty which with the b value is determined. For this purpose we look the average number of earthquakes used to calculate all single values of b in the input array. A detailed explanation of who the b value is calculated can be found in \ref{fig:b_calculation_cylinders}. 
If the number is low, the b values will be calculated on only a few events, which leads to drastic changes between neighboring cells in the input data (even after the autoencoder is used), if a single earthquake is added or removed. Theoretically, better b values will correspond to a better prediction, however better b values will also correspond to either a higher seismicity region, or a region better covered by seismic stations. However, the quality is not given to the model directly, is would have to be determined from the input data.
The impact of the quality is shown in the bottom two panels in Figure \ref{fig:Res_Acc_triple}. The forecasting is clearly dependent on the quality of the data, however, a higher quality does not translate to a higher model output, which is good, since a higher quality does relate to a higher general seismicity. That fact that this does not lead to an immediate higher model output shows the network learns more than just a seimicity estimate.


\subsection{Location Dependence}
\begin{figure}[H]
    \includegraphics[scale=0.8]{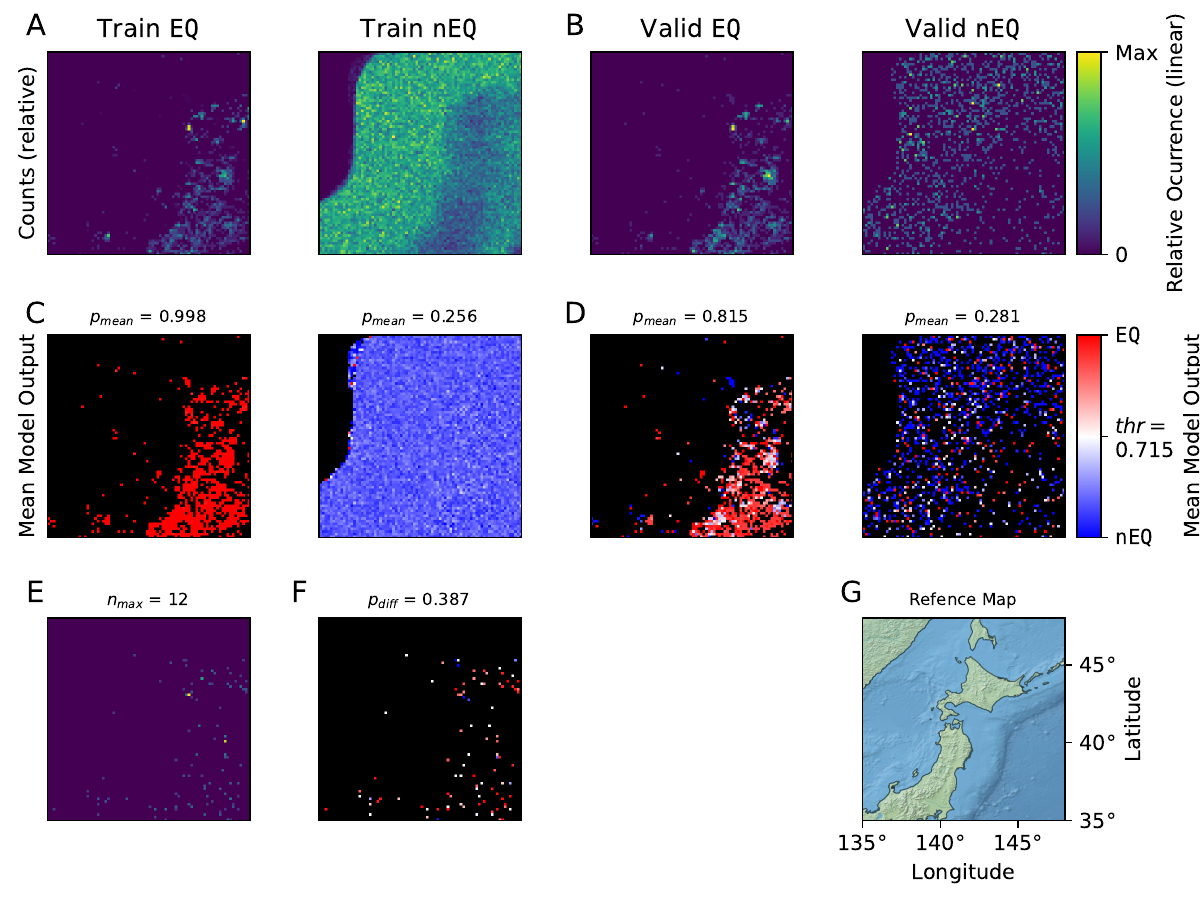}
    \caption{Location related data for Training and Validation datasets. In the top row, the left two panels (A) show where the \EQ and \nEQ class events are selected from for the training dataset, the right to panels (B) show the same for the validation. Note that the validation data of one meta epoch will be in the training data of the next meta epoch. The middle row shows the average model output for the spacial and temporal selections above (C for training, D for validation), the number above the panels is the (count adjusted) average value of the whole panel. The color is scaled to conform to the optimal threshold determined before, while black represents a cell without any events in it. The last row shows from left to right: E) The locations where there are both, \EQ and \nEQ class events in the validation data, as well as their count in that location. F) The difference between the mean model output for the \EQ class and the \nEQ class in the validation data, with the $p_{diff}$, the weighted-by-events-per-cell average of this difference G) A reference map showing the geographical region corresponding to all 11 other panels.}
    \label{fig:Res_Acc_MeanPred}
\end{figure}

\ 
Another possible source of errors is the location. The network could learn where in the domain the sample was taken from and simply predict earthquakes in the regions where there are many earthquakes. While the coordinates of the input are not given, some features can consistently be associated with certain locations, first and foremost the boundary of the area where the b value is calculated (this consistency can be seen in Figure \ref{fig:b_calculation_cylinders}). Figure \ref{fig:Res_Acc_MeanPred} shows the relevant data to determine this has a slight impact: Figure \ref{fig:Res_Acc_MeanPred} A shows the locations for the Training data sets, Figure \ref{fig:Res_Acc_MeanPred} B the locations for for validation sets.
Figure \ref{fig:Res_Acc_MeanPred} C and D show the mean model output for Training and Validation instances respectively. Figure \ref{fig:Res_Acc_MeanPred} C therefore shows very strong overfitting, but Figure \ref{fig:Res_Acc_MeanPred} D shows a clear pattern of accurate forecasting. The number above each of these plots is the weighted mean value in the image. Figure \ref{fig:Res_Acc_MeanPred} E and F illustrate the locations that are included in the validation and training set (at different times), and show.

\subsection{Reproducibility}
During the training process we noticed a heavy dependence on the initial conditions, even to the point of some models being unusable for some initial conditions. This can be seen for the DRN Network in Figure \ref{fig:Architecture_Comparison}. Since a random initialization of the network is common practice, we conducted a study of different seeds to survey this behaviour and find a good seed to work with. For this, we iterated over a fixed seeds for \texttt{numpy} and \texttt{TensorFlow}. We trained a total of 10 model and chose the best one to conduct the analysis presented in this paper. The results of all of them are shown in the supplementary material, see Figure \ref{fig:Res_Seed_Study_AE} A.

\textbf{Impact of the Autoencoder}
We repeated the same reproducibility study, only omitting the autoencoder, using the b value directly as an input to the classifier network and repeating this ten times. With the autoencoder the average accuracy is $68.6\%$ with the best accuracy $72.3\%$, without the autoencoder the average accuracy is $65.0\%$ with $68.7\%$ as the best accuracy. The autoencoder therefore improves the performance by roughly $3.6~\%$. The idea behind using the autoencoder is, that it learns the normal b value images, and that by using the difference the abnormality of the system is highlighted.
The results of both sets of 10 models are shown in Figure \ref{fig:Res_Seed_Study_AE}.

\begin{figure}[H]
    \includegraphics[scale=1]{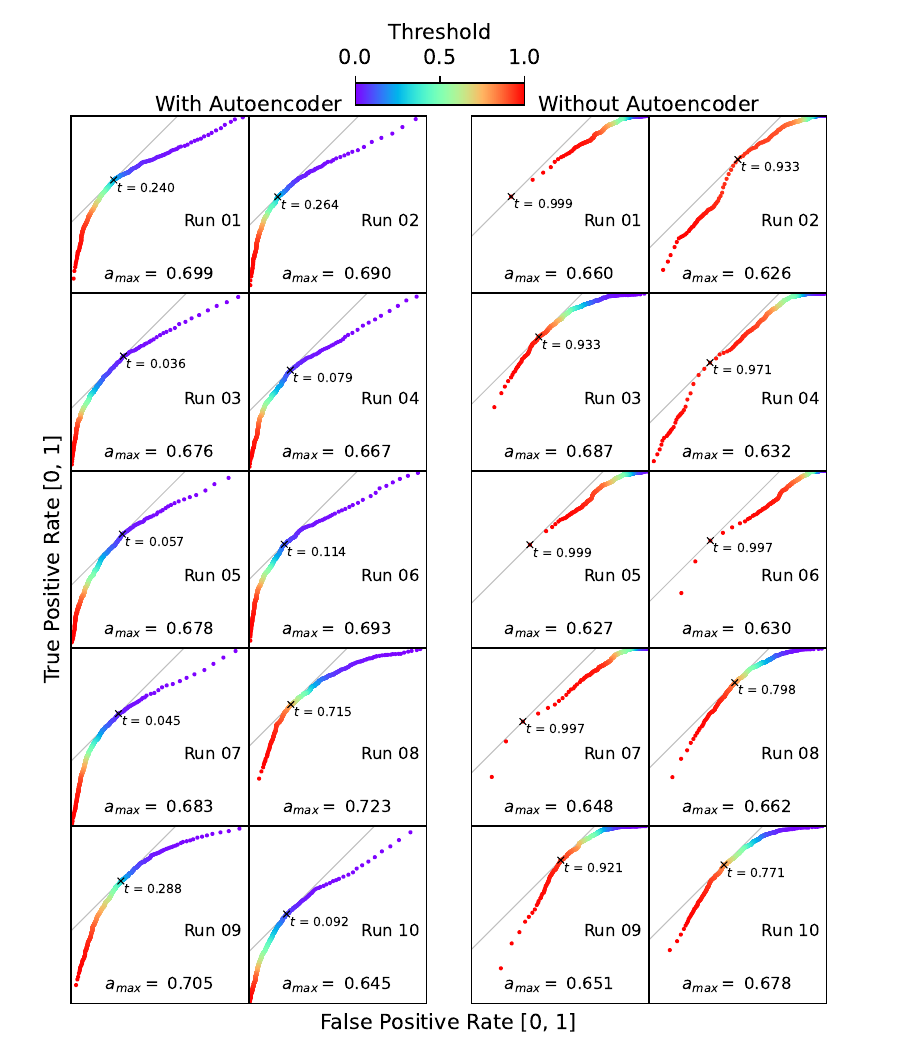}
    \caption{This figure shows the ROC curve for the comparative runs with and without the autoencoder for all 10 different seeds for numpy and tensorflow. The autoencoder was used and the input to the classifier is $\mathrm{AE(data)} - \mathrm{data}$. For the autoencoderless runs, the b value was clipped between 0 and 2.}
    \label{fig:Res_Seed_Study_AE}
\end{figure}

\subsection{Comparison to other work}
There seems to be some possibility in using b value distributions to forecast earthquakes \parencite{Smith1981, Main1989, Smyth2011, Gulia2019}. 
While the aforementioned publications worked by manually (or by the use of classical statistics) looking for patterns in the data, this process is the core ability of deep learning, which makes it a promising approach. This shifts the problem of earthquake forecasting to one of data selection and preparation, where the pattern recognition is left to a deep learning network.

\subsection{Some Caveats}
Concerning the results shown in Figures \ref{fig:Res_Acc_PositiveAnalysis} and \ref{fig:Res_Acc_PositiveAnalysis_all}, the model has the tendency to give long stretches of  high outputs for high seismicity regions, however it does not always give a high output for these regions. There are several interpretations of this, the most charitable being, that the occurrence of the large earthquake is ``ready'' at some point (leading to high model output) and will stay there until the tension is released at a random point in time. This could also distribute the high model output spatially, as \Mw $> 7$ earthquakes have large rupture areas, even exceeding $1000~\mathrm{km}$ for \Mw $> 9$\parencite{Thingbaijam2017}. Following this idea, the model output does not decrease immediately after a large earthquake, because there is still tension, whether new through the earthquake, or retained from before, and it will take a while to release it.
Less charitable would be the interpretation that the model just always gives a high output in high seismicity regions, but this does not seem to be the case, e.g.\ earthquake D in Figure \ref{fig:Res_EXE_Local_Prediction}, but also the low model output times at the other locations -- the earthquakes are in regions that should always be considered high seismicity.

A problem with this dataset is the Tōhoku Earthquake in the temporal center of the data (day 3576 of 6793). An earthquake of that magnitude has the potential to change the seismic behaviour of a region, and it produces a long sequence of significant aftershocks which must be part of a forecast (because they are big enough to be destructive themselves) but are so numerous that they can skew the results in a Machine Learning setting. The impact is visible in all temporal figures presented in this paper. This also has an impact on the previous point of the consistently high model outputs: Since the aftershock sequence is part of the forecast, a high output is somewhat desirable and necessary, if there are aftershocks, even though it would clearly be better to actually have a low model output in between the aftershocks.

Apart from the data considerations, there is also a noticeable variability in the results even with smaller changes in the architecture ($4-5\%$). Even though many such changes were made in the process of this work, it is still likely that a better architecture, even of similar complexity, could be found.
Especially since there many choices during this workflow that can with good reason be done different ways,  optimizing the combination of all those choices can likely improve the overall performance considerably.

\section{Conclusion}

While there are obvious challenges in our model, such as the location dependent overfitting, these do not account for all of the models forecasting performance.
Using only roughly 2300 Earthquakes to train the model shows, that with growing the dataset in the future, better forecasts should be possible. This work also shows the importance of improving the seismic network coverage in earthquake prone regions around the pacific ring of fire where the coverage, and consequently the magnitude of completeness is poorer, for example in Indonesia, Kamchatka, and the west coast of South America.

While our results are quite promising for advancing the problem of earthquake forecasting, there are problems with the short time frame of the training data as well as validation against a fair baseline assumption remain.
Training on more regions which don't have extreme seismicity changing events and including more events from the more recent past as well as changing starting days will very likely improve the models performance in the future.

\newpage
\section{Acknowledgements}
This research is supported by the “KI-Nachwuchswissenschaftlerinnen" -- grant SAI 01IS20059 by the Bundesministerium für Bildung und Forschung -- BMBF. The calculations were performed at the Frankfurt Institute for Advanced Studies’  GPU cluster, funded by BMBF for the project Seismologie und Artifizielle Intelligenz (SAI). We thank Megha Chakraborty, Darius Fenner, Dr.~Claudia Quinteros, Professor Geoffrey Fox and Dr.~Kiran Thingbaijam for their helpful discussion. 
We acknowledge the help and advice from Prof.~Dr.~Horst Stoecker.
The research has made extensive use of TensorFlow \parencite{Tensorflow2015}, numpy \parencite{Numpy2020}, and matplotlib \parencite{Matplotlib2007}.
Geographical maps were made with Natural Earth. Free vector and raster map data @ naturalearthdata.com.
The training was carried out on Nvidia A100 Tensor Core GPUs.

\printbibliography

\end{document}